\documentclass{elsart}

 \usepackage{graphics}
 \usepackage{graphicx}
 \usepackage{epsfig}
 \usepackage{subfigure}

\usepackage{amssymb}

\begin{document}

\begin{frontmatter}



\title{Scaling of the propagation of epidemics in a system
of mobile agents\thanksref{label1}}
\thanks[label1]{In honor to Per Bak.}


\author{M.C. Gonz\'alez\corauthref{cor1}}
\corauth[cor1]{Tel.: +49-711-685-3594; fax: +49-711-6853658.}\ead{marta@ica1.uni-stuttgart.de}
\and\author{H. J. Herrmann}

\address{Institute for Computer Applications 1 (ICA1), University of 
Stuttgart, Pfaffenwaldring 27, 70569 Stuttgart, Germany}

\begin{abstract}
For a two-dimensional system of agents modeled by molecular dynamics,
we si\-mulate  epidemics spreading, which was recently studied on 
complex networks.
Our resulting network model is time-evolving.
We study the transitions to spreading as function of density,
temperature and infection time. In addition, we analyze the epidemic
threshold associated to a power-law distribution of infection
times. 
\end{abstract}
\begin{keyword}
Non-equilibrium phase transitions \sep Contact process \sep Complex networks
\sep Epidemic Dynamics 
\PACS 
\end{keyword}
\end{frontmatter}

\section{Introduction} 
  The statistical spreading of infections, information or damage, 
involves non-equilibrium phenomena. Fluctuations and spatial correlations play an important role and are often not solvable exactly.
Usually these processes are studied on a lattice that can be regular \cite{hirinchsen,Dickman}, hierarchical or small-world \cite{satorras}. But in most cases the population in question is mobile. Therefore, in this work we study a system 
of particles moving according to a simple Newtonian dynamics.
We simulate on it a known contact process \cite{Mollison}, described in terms of a 'SIS' model
of infection, or infection without immunization, that is the state of the 
particles are healthy or infected, and are susceptible to re-infection after 
healing, thus the name of the model (SIS: susceptible-infected-susceptible).\\
We characterize the transition  to spreading of the epidemic dynamics, 
and obtain a continuous range of critical
exponents changing the density of the system. The observed behavior 
results to be  the 'SIS' analogous of a model of 'stirred percolation'
\cite{deGennes,Kerstein,Hede} which was used to describe epidemic 
dynamics with immunization ('SIR').     
\section{Model}
The simulations are carried out on a square shaped cell of
linear size $L$ with periodic boundary conditions. $L=\sqrt{\rho/N}$,
is given by the number of particles ($N$) and the density ($\rho$).
The particles are represented by 'soft-disks' of radius $r_{0}$ moving
continuously on the plane. The interaction between two particles at positions
$\mbox{\boldmath $r$}_i$ and $\mbox{\boldmath $r$}_j$ corresponds to a Lennard-Jones potential truncated at its minimum,
\begin{equation}
u(\mbox{\boldmath $r$}_i,\mbox{\boldmath $r$}_j)= U_{0}\left[\left(\frac{2r_{0}}
{|\mbox{\boldmath$r$}_i-\mbox{\boldmath $r$}_j|}\right)^{12}-2\left(\frac{2r_{0}}{|\mbox{\boldmath $r$}_i-\mbox{\boldmath $r$}_j|}\right)^{6}\right]
+U_{0},\hspace{.05 in}  |\mbox{\boldmath $r$}_i-\mbox{\boldmath $r$}_j|\leq 2r_{0},
\end{equation}
reduced units are used in which $U_{0}$, $r_{0}$ , $k_{B}$ 
(the Boltzmann constant) and $m$ (the particle mass) are all unity.\\    
Along this paper, the particles are considered to be 'agents', to  model 
the 'SIS' process described above. In a simple version one can
assume that each time two agents $(i,j)$ interact or 'collide' (that is, if $|\mbox{\boldmath $r$}_{i}- \mbox{\boldmath $r$}_{j}| \leq 2r_{0}$), the infection propagates from an infected agent to a
susceptible one. In most of our simulations we used a simple initial
condition: (i) at time $t=0$, $N$ agents are distributed regularly in the cell, (ii) have the same absolute velocity $v$ with randomly distributed directions,
(iii) a central agent is infected and the rest are susceptible. At each time
step the positions $\{ {\bf r_{i}} \}$, velocities $\{ {\bf v_{i}} \}$ and the infection state $\{ \sigma_{i} \}$ of the system are updated. We use
the molecular dynamics ($MD$) scheme of cell subdivision with the leapfrog integration method \cite{Rapaport}. 
Once an agent is infected, it heals and becomes 
susceptible again after a fixed number
of time steps, that is the 'time of the infection' ($\Delta t_{inf}$),
and is a free parameter of the model. 
  
\section{Scaling}  
The dynamics of an epidemics is described in terms of the infection rate ($\lambda$). That is defined as the number of agents one agent infects
before healing. Therefore, in this model
\begin{equation}
\lambda \equiv \Delta t_{inf}/\tau_{coll},
\end{equation}
where $\tau_{coll}$ is the mean time between two collisions, and
depends on the mean velocity of
agents ($\langle v \rangle$).
Neglecting the interaction potential with respect to the
kinetic energy, one has:
\begin{equation}
\langle v \rangle = \sqrt{\frac{k_{B} T \pi}{m}}.
\end{equation}
Thus, the mean number of collisions ($\langle n_{coll} \rangle$) of one agent during
a period of time t, is given by
the area within which it interacts ($2r_{0} \langle v \rangle$t),
multiplied by the density,
\begin{equation}
\langle n_{coll} \rangle= \rho \hspace{.05 in} 2r_{0} \langle v \rangle t.
\label{eq:coll1p}
\end{equation}
The pre-factor of time in eq.~(\ref{eq:coll1p}) is the collision frequency,
and its inverse, gives the mean time between collisions, 
\begin{equation}
\tau_{coll} = \frac{1}{ \rho \hspace{.05 in} 2r_{0}} \sqrt{\frac{m}{\pi T k_{B}}}
\label{eq:factor}
\end{equation}

\begin{figure}
\centering
\includegraphics[width=10.5cm]{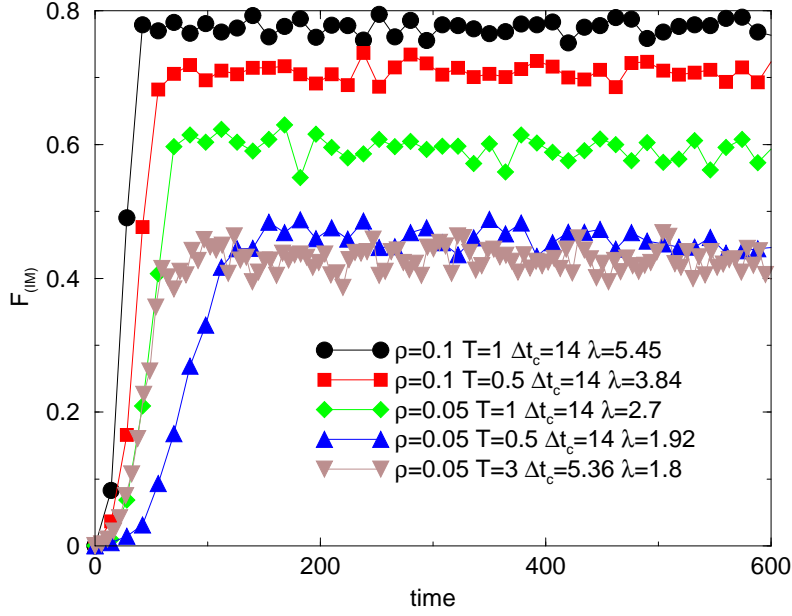}
\caption{Starting with one infected agent, the figures shows how the
infection spreads with time and after a certain 
period of time ('transient'), the fraction of
infected agents ($F_{IM}$) fluctuates around a mean value. The fraction of 
infected agents in this 'quasi-equilibrium' state, increases with the 
infection rate ($\lambda$).}
\label{fig:Fvst}
\end{figure}

\begin{figure}
\centering
\includegraphics[width=10.5cm]{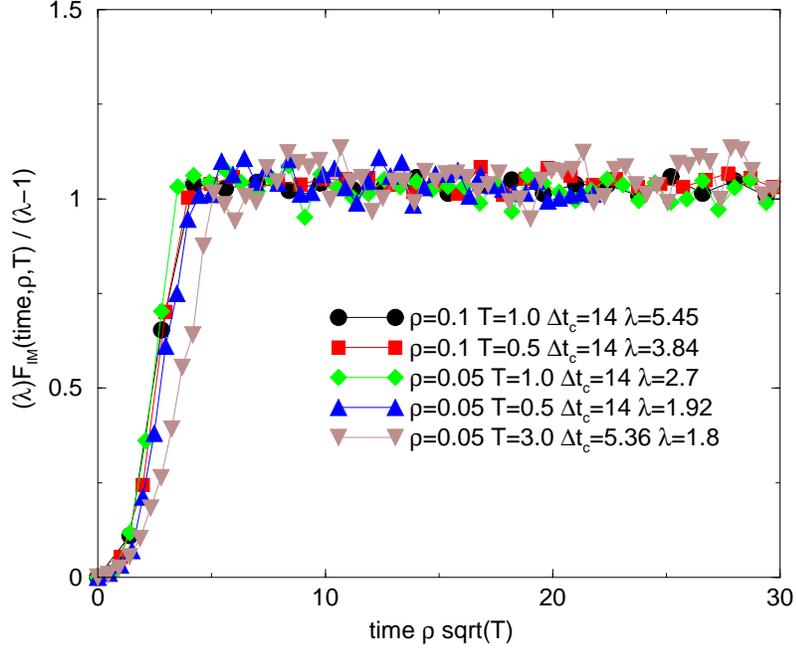}
\caption{The plot shows the collapse of data of
Fig.~\ref{fig:Fvst}. The vertical axis was divided by the mean
field approximation of $F_{IM}(\lambda)$ and the horizontal axis
is divided by the collision time ($\tau_{coll}$).}
\label{fig:collapses}
\end{figure}

In Fig.~\ref{fig:Fvst} we see the fraction of infected 
agents ($F_{IM}(t)$ vs. $time$) for 
different values of temperature ($T$), density ($\rho$) and 
time of infection ($\Delta t_{inf}$). In each case we start with one 
infected agent and after a transient, the system
fluctuates around a value of $F_{IM}$, which
depends on $\lambda$. This mean value can be calculated
using a mean field approach,
\begin{equation}
\frac{\partial F_{IM}(t)}{\partial t} = -F_{IM}(t) + \lambda F_{IM}\left[1-F_{IM}\right]    
\end{equation} 
The first term of the r.h.s is the fraction of agents that heals and the
second term, is the fraction of agents that becomes infected.
After the transient, $\partial F_{IM}(t)/\partial t \sim 0$. Thus
\begin{equation}
                F_{IM}(\lambda)       = \left\{ \begin{array}{ll}
                                     0     & \mbox{if $\lambda \leq \lambda_{c}$}\\
                                     1-1/\lambda   & \mbox{if $ \lambda > \lambda_{c}$}
                                     \end{array}
                               \right.
\label{eq:MF} 
\end{equation}
where $\lambda_{c}=1$, is known as the critical rate of infection. 
In Fig.~\ref{fig:collapses} we see the collapse of 
the realizations of Fig.~\ref{fig:Fvst}, with the expression
obtained in eq.~(\ref{eq:MF}). The dynamics of the system
is characterized by the infection rate $\lambda$, which contains 
all the free parameters of the system.\\

\section{Transition to spreading}

\begin{figure}
\centering
\includegraphics[width=10.5cm]{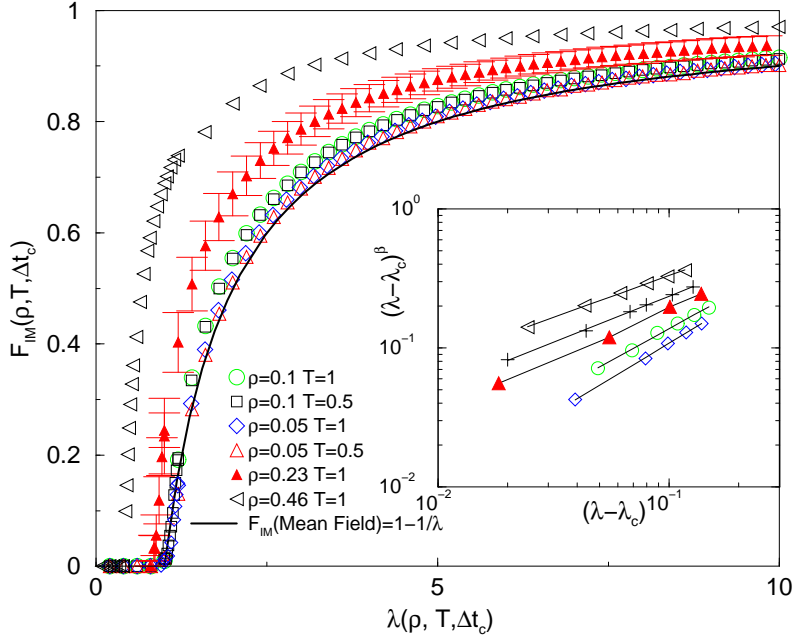}
\caption{Fraction of infected nodes $F_{IM}$ as function of $\lambda$
for different conditions of $T$ and $\rho$.
The data agree with the $MF$ approximation
(eq.~(\ref{eq:MF})) only for low densities ($< 0.1$). 
The inset shows the $\log-\log$ plot of $F_{IM}$
vs. $\lambda - \lambda_{c}$ with $T=1$ and increasing densities: 
$\rho=0.05$, $\lambda_{c}=1.05$ (diamonds),
$\rho=0.1$, $\lambda_{c}=1.06$ (circles), $\rho=0.23$, $\lambda_{c}=0.862$ (filled triangles), $\rho=0.30$, $\lambda_{c}=0.73$ (crosses),
and $\rho=0.46$, $\lambda_{c}=1.05$ (rotated triangles). The lines are regressions of the data and have respectively slopes: $\beta=1.0$, $\beta=0.92$, $\beta=0.738$, $\beta=0.66$, and $\beta=0.599$.}
\label{fig:FIM}
\end{figure}

\begin{figure}
\centering
\includegraphics[width=10.5cm]{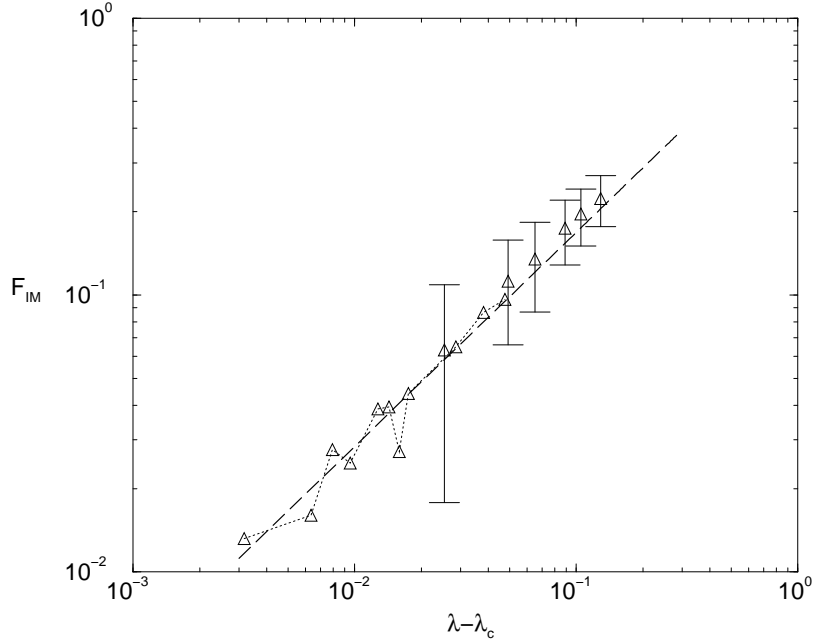}
\caption{$Log-log$ plot of the fraction of infected nodes $F_{IM}$
as function of $\lambda - \lambda_{c}$ for ($\rho=0.2$, $T=1$),
with $\lambda_{c}=0.935$. The dashed line is a fit to
the form $F_{IM} \sim (\lambda - \lambda_{c})^{\beta}$, with 
exponent $\beta=0.773$.}
\label{fig:02beta}
\end{figure}

We analyze the transition to spreading of the epidemic. 
In systems of $1254$ agents 
$F_{IM}$ after      
the transient is averaged, over $500$ initial conditions
at a given $\lambda$.
In Fig.~\ref{fig:FIM} we see that the results agree
with the mean field approximation for densities
lower than $0.1$. Increasing the density, changes
the shape of the transition curve. Near the critical point
the $F_{IM}$ follows a power law, 
$F_{IM}\sim (\lambda - \lambda_{c})^{\beta}$.
The inset of Fig.~\ref{fig:FIM} in a double-logarithmic plot,
demonstrates this power law behavior, with straight lines
of slope $\beta$. As can be seen, the value of $\beta$
depends on the density of the system. Smaller values of $\beta$
indicate more significant changes of $F_{IM}(\lambda)$
near the transition. Even for low densities
(i.e. $\rho \in [0.1,0.2]$), where one expects
$MF$ to be valid, the critical exponents change
with the density.\\
 In order to confirm this observation, we study
the transition in detail. Fig.~\ref{fig:02beta} 
shows  the $\log-\log$ of 
$F_{IM}$ vs. $(\lambda -\lambda_{c}$), averaging
over $1000$ realizations in  systems with $\rho=0.2$, $T=1$, and
$\lambda_{c}=0.935$. We see that the data fit the expression 
$\sim (\lambda -\lambda_{c})^{\beta}$, with exponent $\beta=0.773$.\\

\begin{figure}
\centering
\includegraphics[width=11.5cm]{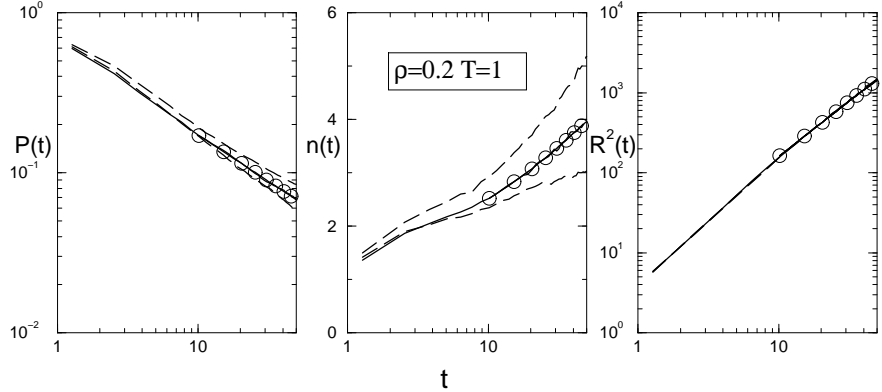}
\caption{Evolution of the survival probability of infection $P(t)$, the 
mean number of infected agents $n(t)$ and the mean square distance of 
spreading $R^{2}(t)$ in time. Each graph contains three curves near 
criticality. With $\rho=0.2$, from bottom to top: $\lambda=0.93$, $0.94$ 
and $0.95$. The circles come from regressions to calculate the critical 
exponents}
\label{fig:Ptnt}
\end{figure}

Other critical exponents, are obtained, if
one averages the survival probability of infection $P(t)$, the number 
of infected agents $n(t)$ and the square distance of spreading $R^{2}(t)$ 
and plots them against time. At the critical point, they are expected to display asymptotic power laws 
\cite{Grassberger},
\begin{eqnarray}
P(t) \sim t^{-\delta}, \hspace{.3in} n(t) \sim t^{\eta}, \hspace{.3in} R^{2}(t) \sim t^{z}.
\label{eq:critic}
\end{eqnarray} 
The relations in eq.~(\ref{eq:critic}) apply at long times, and require
that the infection does not reach the boundaries of the system.
Results for the three quantities $P(t)$, $n(t)$ and $R^{2}(t)$, averaged
over $\sim 10^{3}$ realizations, with systems of $\sim 10^{4}$ agents, and 
fixed temperature ($T=1$), are shown in Fig.~\ref{fig:Ptnt} for 
$\rho =0.2$.

\begin{table}
\begin{center}

\begin{tabular}{ccccc} \hline\hline
             &$MF$ & $\rho=0.1$ & $\rho=0.2$ & \mbox{Contact Process(2D)\cite{Dickman}} \\
$\lambda_c$  &  1   &    1.0(8)    &    0.9(4)    &      1.6488(1)\\
$\beta$      &  1   &    0.9(2)    &    0.7(7)    &      0.583(4)\\
$\delta$     &  1   &    0.5(9)    &    0.5(3)    &      0.4505(10)\\
$\eta$       &  0   &    0.1(5)    &    0.2(5)    &      0.2295(10)\\
$z$          &  1   &    1.3(0)    &    1.2(7)    &      1.1325(10)\\
\hline \hline
\end{tabular}
\end{center}
\caption{Critical rate of spreading and exponents 
for $SIS$ model on moving agents 
with $\rho=0.1$ and $0.2$, contact process in two dimensions, 
and estimates obtained by mean field $MF$\label{table:ec}.}
\end{table}

\begin{figure}
\centering
\includegraphics[width=8.5cm]{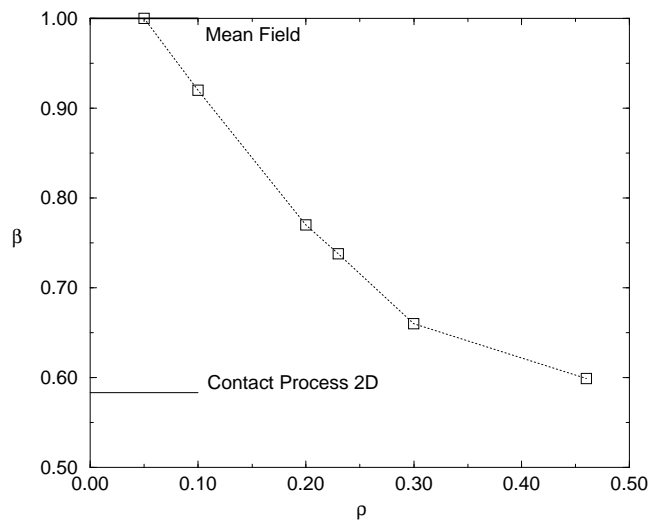}
\caption{Numerical estimates for the critical exponent
$\beta$, for systems with different density.}
\label{fig:beta}
\end{figure}

The values of the critical exponents are reported in table~\ref{table:ec}.
The known hyper-scaling relation of the dynamics
$4\delta+2\eta=dz$, where $d$ is the dimension of the system,
is recovered within the range of numerical errors.
Increasing the density of the system gives a continuous change 
in the critical exponents of the epidemic dynamics, they
go from $MF$ values to the exponents for contact process in 
a two dimensional lattice \cite{Dickman}. Numerical estimates of the critical
exponent $\beta$ vs. $\rho$ are shown in Fig.~\ref{fig:beta}.\\

The dependence of the critical exponents of the dynamics on the density of
agents ($\rho$), is analogous to the dependence on the 'flammable' 
fraction of space ($\phi$), observed in 'forest fire' dynamics, 
or epidemics dynamics with immunization
\cite{Kerstein,Hede}, described through 'stirred percolation' 
models. 'Stirred percolation' consists in random walkers on a 
changing lattice and was proposed by de 
Gennes~\cite{deGennes} to describe the behavior
of conductivity in binary mixtures \cite{Lagues}.                                                                                

\section{Power-law distribution of infection times}
\begin{figure}
\includegraphics[width=12.5cm]{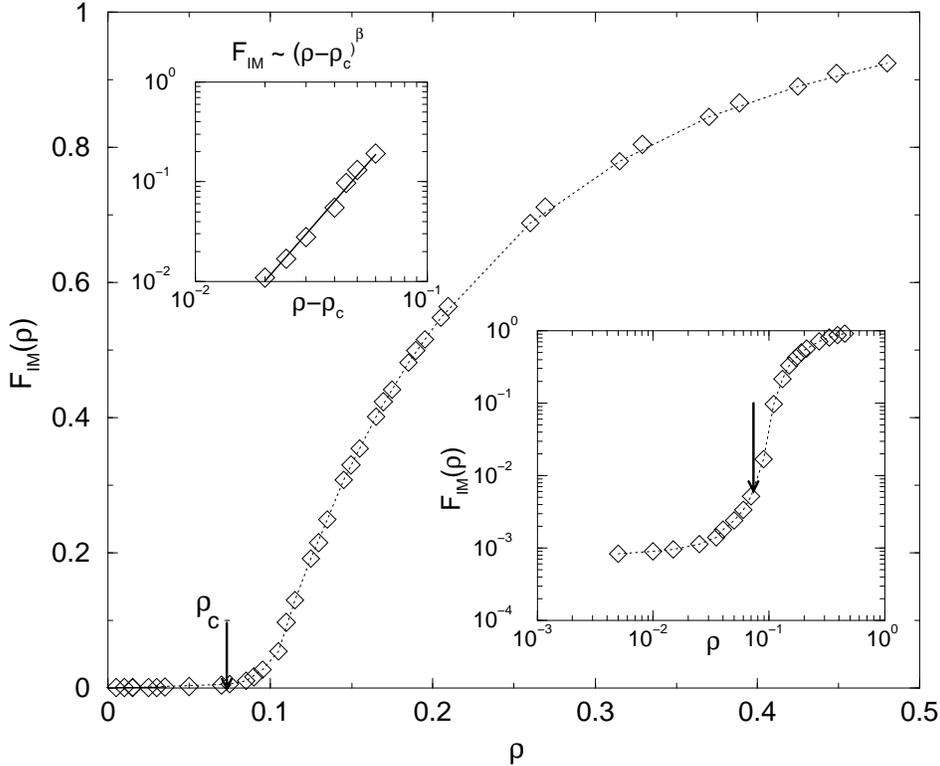}
\caption{Fraction of infected nodes vs. density, with power-law
distribution of infection time (eq.~\ref{eq:PL}) for $\gamma=2.4$.
$F_{IM} \sim (\rho-\rho_{c})^{\beta}$ is shown in the upper-left-corner 
inset  with $\rho_{c}=0.065$ and $\beta=2.6(5)$. The bottom-right-corner
inset is the main plot in $\log-\log$ scale to see better the tail of
the spreading curve, for $\rho<\rho_{c}$.}
\label{fig:rho_c}
\end{figure}

Considering the same  value of infection times ($\Delta t_{inf}$)
for each agent co\-rres\-ponds to situations with homogeneous connectivity. 
In order to extent the model to some real world situations where 
the number of contacts varies greatly from one agent to another, we assign
$\Delta t_{inf}$ to each agent following a power-law distribution,
\begin{equation}
P(\Delta t_{inf}) = (\gamma-1)\Delta t_{inf}^{-\gamma} \hspace{0.5 in} \Delta t_{inf} \geq 1.
\label{eq:PL}
\end{equation}
As a result, for $2<\gamma\leq 3$, the epidemic threshold tends to 
zero, like has been observed for $SF$ networks~\cite{satorras} (see
Fig. \ref{fig:rho_c}).\\
However, the shape of the spreading curve has
a point of inflection ($\rho_c$) above which, the infection 
is much larger ($F_{IM}\sim(\rho-\rho_{c})^{\beta}$, 
$\rho_c=0.06(5)$, $\beta=2.65$, for $\gamma=2.4$).\\
We see that the more 'connected'
agents are responsible for the absence of an epidemic threshold
(tail of the curve). The infection would spreads only
if the {\em mean} rate of spreading is larger than one, that is
\begin{equation}
\lambda_{c} \equiv \frac{\langle \Delta t_{inf} \rangle}{\tau_{coll}} = \frac{\gamma-1}{\gamma-2} \hspace{.05 in} \rho \hspace{.05 in} 2r_{0} \sqrt{\frac{\pi T k_{B}}{m}} > 1. 
\label{eq:mean_lambdac}   
\end{equation}
$\lambda_{c}=1$ gives the inflection point $\rho_{c}$, which for
$\gamma=2.4$ is $\rho_{c}=0.0718$ (according to eq.~\ref
{eq:mean_lambdac}, for $r_{0}=2^{1/6}$ and $T=1$), and agrees with
the numerical value $\rho_{c}=0.06(5)$ reported in Fig.~\ref{fig:rho_c}
averaging over about $10^{3}$ realizations with about $10^{3}$ agents
each.

\section{Conclusions}
Novel effects are observed studying the $SIS$ model of infection
on a system of moving agents. A continuous range of
critical exponents are observed as function of the density
of agents, recovering mean field predictions for lower densities,
two-dimensional exponents of contact process, increasing the
density. Introducing a power-law distribution of infection times,
the epidemic threshold becomes zero due to
the more `infecting agents'; but still there is a {\em critical} 
rate of spreading, which depends on the exponent of 
the distribution and the mean time of interaction among the agents.

\end{document}